\documentclass[10pt,twocolumn]{article}

\usepackage{multicol}
\usepackage{graphicx}
\usepackage{amsmath}
\usepackage{amssymb}
\usepackage{bm}
\usepackage{lipsum}
\usepackage{color}
\usepackage{siunitx}
\usepackage{authblk}
\usepackage[margin=2cm]{geometry}

\title{Deposition of quantum dots in a capillary tube}
\author[1]{Yong Lin Kong}
\author[1]{Fran\c{c}ois Boulogne}
\author[1]{Hyoungsoo Kim}
\author[1]{Janine Nunes}
\author[1]{Jie Feng}
\author[1]{Howard A. Stone}
\affil[1]{Department of Mechanical and Aerospace Engineering, Princeton University, Princeton, NJ 08544}
\date{\small\today}
\begin{document}

% pacs: 82.70.Dd    Colloids
% 87.10.Pq Elasticity theory

\twocolumn[
    \begin{@twocolumnfalse}
        \maketitle
        \begin{abstract}
The ability to assemble nanomaterials, such as quantum dots, enables the creation of functional devices that present unique optical and electronic properties. For instance, light-emitting diodes with exceptional color purity can be printed via the evaporative-driven assembly of quantum dots. Nevertheless, current studies of the colloidal deposition of quantum dots have been limited to the surfaces of a planar substrate. Here, we investigate the evaporation-driven assembly of quantum dots inside a confined cylindrical geometry. Specifically, we observe distinct deposition patterns, such as banding structures along the length of a capillary tube. Such coating behavior can be influenced by the evaporation speed as well as the concentration of quantum dots. Understanding the factors governing the coating process can provide a means to control the assembly of quantum dots inside a capillary tube, ultimately enabling the creation of novel photonic devices.
        \end{abstract}
    \end{@twocolumnfalse}
]

\section{Introduction}
Nanomaterials enable the creation of functional devices that possess unique properties. For instance, cadmium selenide-zinc sulfide quantum dots (QDs) \cite{Alivisatos1996} have a narrow emission spectra that enable the creation of light-emitting diodes with exceptional color purity \cite{Shirasaki2013}. Further, such nanomaterials can be dispersed into solvent to form solution-processable inks, which can be integrated with coating or printing processes to create photonic devices on a two-dimensional substrate \cite{Wood2009} or on a three-dimensional construct \cite{Kong2014}. The integration of electronics with a variety of constructs could enable new applications, such as the realization of next-generation biomedical devices \cite{Kim2010} or on-body electronics \cite{Kim2011c}. However, most of the studies focus on the coating of nanomaterials onto an open surface. Yet, in a three-dimensional configuration, the total area of a confined surface can be significantly larger than its  exposed surface. More importantly, in many biologically evolved or synthetically engineered systems, such as the transport of oxygen and nutrient in a blood vessel or a DNA analysis in a microfluidic chip, many of the critical  processes occur inside a confined region rather than on its open surface. Hence, the ability to coat the surface within a confined space with nanomaterials could enable the integration of functional devices for new applications.

Unfortunately, the confined nature prohibits the coating of such surfaces by using conventional lithography \cite{Coe2002} or many of the novel processing strategies that were developed for open surfaces such as  dip-pen nanolithography \cite{Piner1999a}, transfer \cite{Ahn2006}, or 3D printing \cite{Kong2014}.  On the other hand, evaporative self-assembly is a promising method in which the nanoparticles can be assembled into a variety of patterns by manipulating parameters such as the evaporation rate  \cite{Bigioni2006}, Marangoni effects, \cite{Hu2006} etc. For instance,  on a planar surface, the self assembly of micrometer size band structures have been observed in the literature \cite{Adachi1995, Xu2007, Watanabe2009,Bodiguel2010, Kim2010c, noguera2014transition}. Such patterns are believed to originate from the pinning of the contact line of a droplet during the evaporation process, which causes the accumulation of the particles near the contact line in a so-called coffee ring effect  \cite{Deegan1997,Deegan2000}. As the suspension evaporates, the contact line depins before it is pinned again at the next position, and in this sequential way, stripes or multiple irregular rings are formed  \cite{Shmuylovich2002}. 

In contrast, experiments that involve the active retraction of a substrate or meniscus may be described as an active set-up \cite{Thiele2014}, and have been shown to be able to generate highly regular bands in a planar configuration. The withdrawal speed of the substrate from a liquid bath of colloidal solution can be controlled to create a highly regular array of band structures \cite{Watanabe2009}. The formation of bands are described as a consequence of the breakage of the meniscus as it is being stretched during the evaporation \cite{Watanabe2009}.The substrate withdrawal speed in such configurations plays a critical role in determining the coating behavior, which can typically be classified into an evaporative regime for small capillary number and Landau-Levich regime for larger capillary number, where the capillary number represents the ratio of viscous to interfacial stresses. \cite{doi:10.1021/la803646e,doi:10.1021/la1018373,doi:10.1021/jp9114755,Berteloot2012}. The readers may refer to a recent review article on patterned deposition \cite{Thiele2014} for a comprehensive list of references regarding the coating of open surfaces.

Nevertheless, significantly fewer studies have been conducted to investigate the coating of a confined space. The confinement of a closed space imposes a limitation on the evaporation rate.  Previous studies have been performed in capillary tubes with silica particles \cite{Wang2002,nakamura2004simple,kuo2006colloidal,Bodiguel2010}, polystyrene beads \cite{Abkarian2004,kuo2006colloidal}, gold nanoparticles \cite{kuo2006colloidal}, tobacco mosaic virus \cite{Lin2010a} and proteins \cite{Lin2010b}. Our study differs in three major aspects. First, the majority of the previous studies were done with structurally simple micron-size particles that are spherical in shape. In contrast, typical functional nanoparticles such as QDs have a more complex structure with a polyhedral shape. For instance, (CdSe)ZnS QD is a core-shell particle that consists of a small inorganic semiconductor core and a wider-bandgap inorganic semiconductor shell that is passivated with organic ligands such as octadecylamine. The organic ligands confer solubility for the QDs in a non-polar solvent, but they also play an important role that affects the self-assembly during the deposition process \cite{LekkerkerkerHenkN.W.Tuinier2011}. Secondly, most of the previous studies in a confined geometry have not examined the real-time dynamics of the deposition. In our study, such observations are possible as the intrinsic photoluminescence with narrow width of emission of QDs enables the real-time visualization of the nano-particle coating process without the need to tag with fluorescent polymers, which potentially affects the deposition behavior. Further, QDs have been shown to exhibit fluorescence lifetimes that are 10 - 100 times longer than organic fluorescent polymers \cite{resch2008quantum}. This enable reliable visualization and consistent measurement of the coated QDs pattern based on the photoluminescence intensity. Third, the ability of visualizing the particles enables the study of nanoparticle coatings at significantly lower concentrations than previous studies. The volume fractions that have been studied here are approximately 10 - 100 times (0.02 - 0.4 vol \%) lower than the range that was typically studied in other reports in the literature. This limit is important for the purpose of nanoelectronic printing where a thin layer deposition of nanoparticles is required \cite{Shirasaki2013,Kong2014}. 

\section{Experimental Section}

\begin{figure}
    \includegraphics[width=\linewidth]{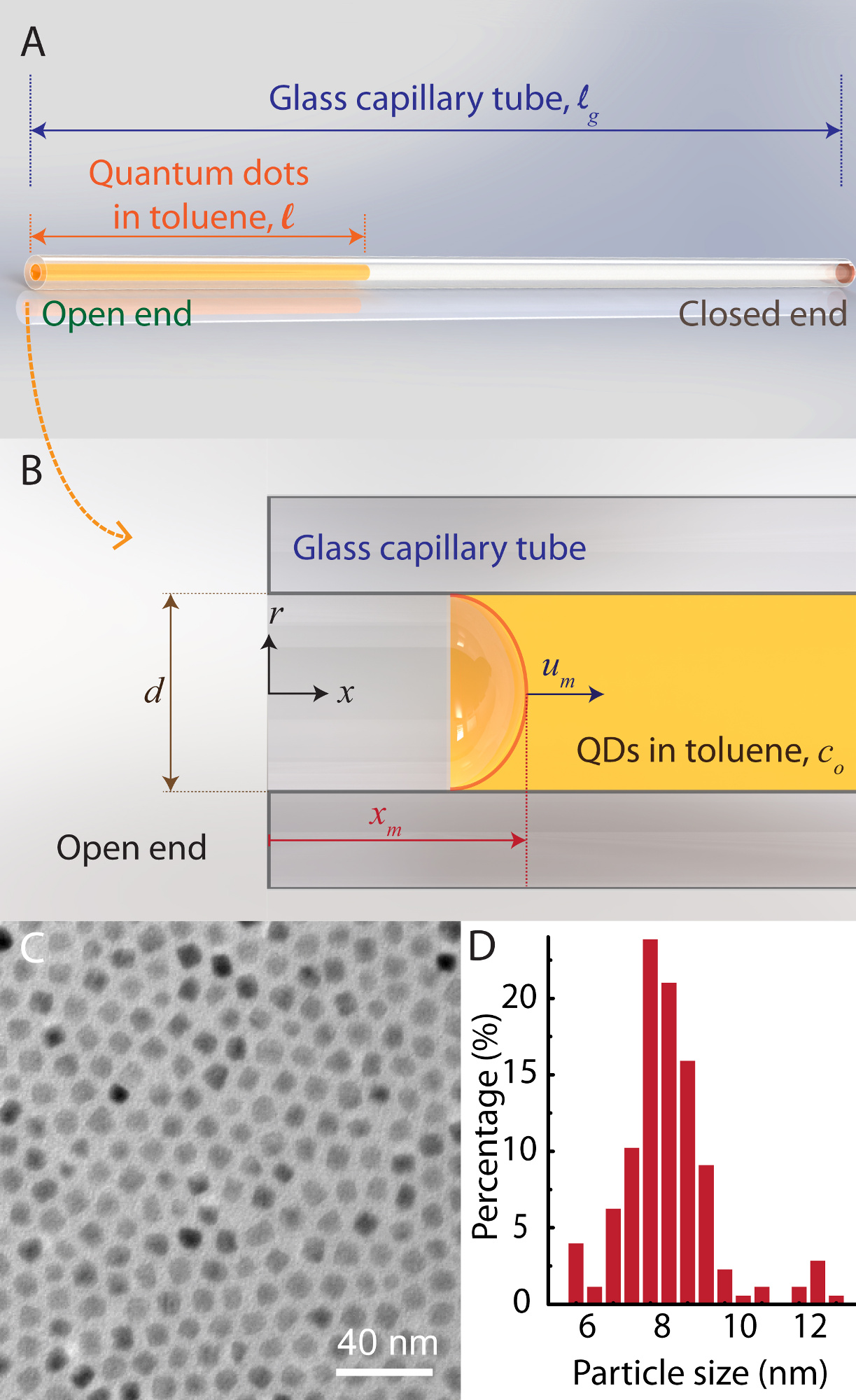}
  \caption{Experimental schematic and CdSe/ZnS quantum dots (QDs). \textbf{(A)} QDs suspended in toluene  are loaded to initial length $\ell$ via capillary forces into a glass capillary tube of length $\ell_g$. The right end of the tube is then sealed and the tube is placed on a horizontal surface with an open end (left side). \textbf{(B)} Cross-section of the tube at the open end. Toluene evaporates from the open end (diameter $d$) and the meniscus moves towards the closed end with velocity $u_m$. The distance between the meniscus and the tube's open end is defined as $x_m$. \textbf{(C)} Transmission electron microscope image of the QDs, which have polygonal shapes in cross section. \textbf{(D)} The QDs have an average size distribution around 8 nm. }
  \label{Fig1}
\end{figure}
The fabrication of functional devices inside a confined geometry requires the controlled deposition of the nanoparticles inside the surface.  Our study aims to understand these coating processes that can ultimately be used to create such functional devices inside a closed construct such as a cylindrical channel. In our experiments, a glass capillary tube (Wiretrol, Drummond Scientific Company, Broomall, PA)  with diameter $d$ and length, $\ell_g$, of 75 mm is first held vertically with a linear stage (PT1, Thorlabs, Newton, NJ)  and immersed into a suspension of QDs. The particles used were cadmium selenide-zinc sulfide QDs (QSP-600, Ocean Nanotech, San Diego, CA). The QDs are dispersed in toluene (Sigma-Aldrich, St. Louis, MO) at concentration $c_0$ from 0.05 mg/ml to 1.00 mg/ml ($\approx$ 0.02 vol\% to 0.40 vol\%). The suspension fills the tube to a length $\ell_0$. The other end is sealed with hot melt adhesive before the capillary tube is carefully withdrawn from the suspension. 

The capillary tube is then placed on a horizontal surface, with a loaded suspension of a length $\ell$, as shown in Figure \ref{Fig1}A. Toluene evaporates from the open end and the meniscus moves towards the closed end with velocity, $u_m$, as sketched in Figure \ref{Fig1}B. The distance between the meniscus and the tube's open end is defined as $x_m$. In contrast, any gas trapped at the closed end is saturated with toluene. Hence, the meniscus at the closed end will have a negligible movement in comparison with the meniscus at the open end. This effect isolates the evaporation area to the meniscus at the open end of the tube. In comparison, previous studies of the colloidal deposition inside capillary tubes have been done by vertically \cite{Abkarian2004,Guo2012} immersing the tube into suspensions and allowing both the bath and the liquid in the tube to evaporate simultaneously, where the movement of the liquid is determined by the rate of evaporation from both the open end of the capillary tube as well as the bath reservoir.  

Here, the samples are illuminated with ultra-violet  light and the images are acquired with a digital single-lens reflex camera (D5200, Nikon, Tokyo, Japan) with a macro lens  (85 mm f/3.5G Micro Nikkor Lens, Nikon, Tokyo, Japan) that is attached to a custom-made stage. The experiments are conducted in a glove box covered with dark cloth to minimize background noise for subsequent image processing. The temperature on the stage of the samples is found to be stable and consistent at 26$\pm$1$^\circ$C throughout the experiments, which is slightly above the room temperature due to the heat from the light source.

The QDs are characterized with transmission electron microscopy (CM100, FEI, Hillsboro, OR). As described above, CdSe/ZnS QDs are core-shell nanoparticles with polygonal shape as shown in Figure \ref{Fig1}C. By measuring the diameter of the particle obtained by assuming circularity, the TEM image shows the mean size distribution of 8 nm in Figure \ref{Fig1}D. In addition to the real-time images acquired with the camera, the deposited QD features are characterized with confocal microscopy (TCS SP5, Leica Microsystems, Wetzlar, Germany). To examine the inner surface, we first cleaved the tubes into smaller sections of approximately 3 mm in length with a diamond scriber. The sectioned tube is then attached to a double sided carbon tape before the capillary tube is scribed along the length of the tube with a diamond scriber. The scribed capillary tube is then cracked open with a metal rod, where the outer surface of the capillary tube pieces would be attached on the double sided carbon tape. A fluorescent image is taken to identify the region of interest prior to imaging with Scanning Electron Microscopy (Quanta 200 FEG ESEM, FEI, Hillsboro, OR), and Atomic Force Microscopy (Dimension NanoMan, Veeco, Plainview, NY).

\section{Results and Discussion}
\begin{figure}
\centering
\includegraphics[width=\linewidth]{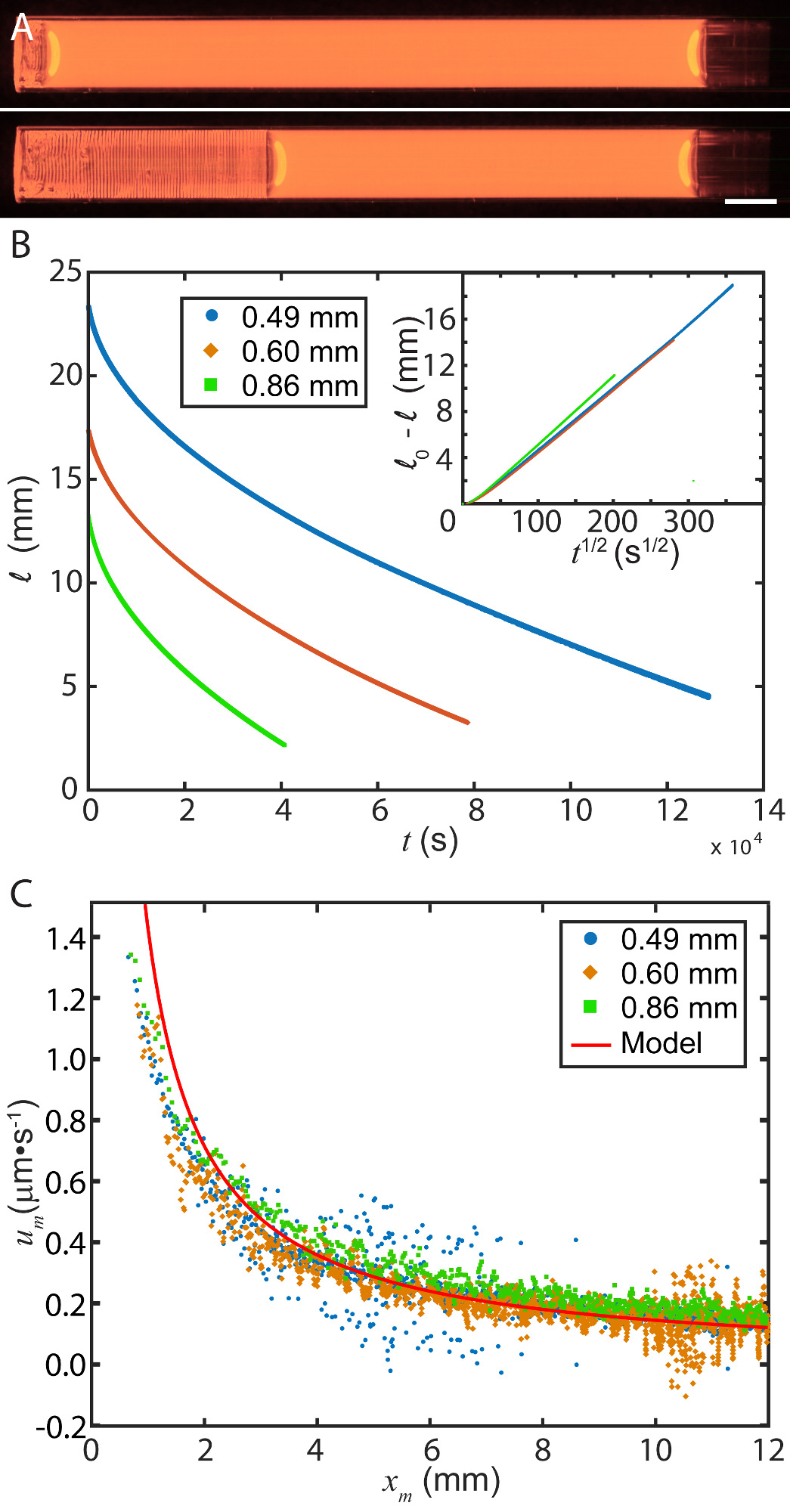}
\caption{Evaporatively-driven motion of the meniscus inside the capillary tubes. \textbf{(A)} Time series images show the column length $\ell$ as a function of time for a 0.86 mm diameter capillary tube (interval = 160 minutes); the scale bar is 1 mm. \textbf{(B)} The column length $\ell$ versus time for different tube diameters, $d$ = {0.49, 0.60 and 0.86} mm, as toluene evaporates; $c_0$ = 0.50 mg/ml. The inset shows the movement of the meniscus follow $t^{1/2}$, where $\ell_0$ is the initial column length. \textbf{(C)} The speed, $u_m$, of the meniscus as it travels along the capillary (\textit{x}); $c_0$ = 0.50 mg/ml. The speed decreases with $x_m$. The evaporation is limited by the diffusion of the toluene from the meniscus front to the opening of the tube, and can be described with a theoretical model derived from Fick's first law without the need of fitting parameters (solid red line).}\label{Fig2}
\end{figure}

When QD suspensions are illuminated with ultra-violet light, the image processing of the photoluminescence intensity enables the tracking of the evaporatively-driven motion of the menisci inside the capillary tubes such as the images shown in Figure \ref{Fig2}A. Figure \ref{Fig2}B shows measurements of the meniscus positions versus time $t$ for three different tube diameters. The amount of toluene lost ($x_m =\ell_0 -$ $\ell$) scales linearly with $t^{1/2}$ as plotted in the inset, which suggests that the evaporation is limited by the diffusion of the toluene molecules to the opening of the tube. We hypothesize that the drying dynamics is controlled by the vapor phase. To examine this hypothesis, we adapted the derivation of  Erbil et. al \cite{Erbil2002} to describe our experimental results, where the system is approximated with Fick's first law. We assume that the P\'eclet number, which represents the balance between convection and diffusion, is small, i.e.   Pe = $\frac{x_m^2}{D_0 t} \ll 1$. Hence, for the vapor phase concentration, $c_v$, this limit corresponds to a quasi-stationary state at the region away from the tube opening such that $\frac{\partial c_v}{\partial t}=0$. 

\begin{equation} \label{eq2}
u_m =  \frac{{\rm d}x_m}{{\rm d} t}= -\frac{D_0}{\rho} \frac{\partial c_{v}}{\partial x},
\end{equation} where $c_v$ is the vapor concentration, $D_0$ is the vapor phase diffusion coefficient ($8.11\times 10^{-6}$ m$^2$/s) \cite{Erbil2002} and $\rho$ is the liquid phase density of toluene ($8.61 \times 10^2$ kg/m$^3$). Further, we can also assume that the toluene vapor concentration is zero at the outlet. We can then approximate
\begin{equation} \label{eq3}
\frac{\partial c_{v}}{\partial x} = \frac{c_{e} - 0}{x_m - 0} =  \frac{c_{e}}{x_m} = \frac{P_{sat}M}{RTx_m},
\end{equation} 
where we have used the ideal gas law with $P_{sat}$, the saturated vapor pressure at the absolute temperature of 299 K ($4.00 \times 10^{3}$ Pa), \textit{M} the molecular weight ($9.21 \times 10^{-2}$ kg/mol), \textit{R} the ideal gas constant (8.31 J/mol K), and \textit{T} the absolute temperature (299 K). Since the vapor pressure of the toluene is significant, we then incorporate Stefan's general diffusion theory  \cite{Marrero1972} as a correction factor, where  $P =  P_{T}\ln \left(\frac{P_T}{P_T - P_{sat}}\right)$ with $P_{T}$ is the total pressure ($1.01\times 10^{5}$ Pa). Combining equations (\ref{eq2}) and (\ref{eq3}) yields
\begin{equation} \label{eq6}
u_m =  \frac{K}{ x_m} ,
\end{equation}
 where $K = \frac{D_0MP_T}{\rho RT} \ln \left(\frac{P_T}{P_T - P_{sat}}\right)$. The integration of equation (\ref{eq6}) yields $\ell(t) = \ell_0 - \sqrt{2Kt}$ where $\ell(t) = \ell_0 -x_m$. This behavior, $u_m \propto x_m^{-1}$, describes the reduction of evaporation speed with distance, $x_m$, as shown in Figure \ref{Fig2}C. We note that the prefactor is independent of the capillary tube diameter, as shown by the solid red line in Figure \ref{Fig2}C.  The prediction (the solid red line in Figure \ref{Fig2}C), which includes the prefactor \textit{K},  matches our experimental data in a region away from the tube entrance where Pe = $\frac{x_m^2}{D_0 t} \approx 10^{-4} \ll 1$. This supports our hypothesis that the evaporation is limited by the diffusion of the evaporated toluene molecules to the opening of the tube. This drying dynamics is different from the confined unidirectional drying with a high volume fraction of particles \cite{Allain1995,wallenstein2011theory,lidon2014dynamics}. 

\begin{figure}
\centering
\includegraphics[width=.8\linewidth]{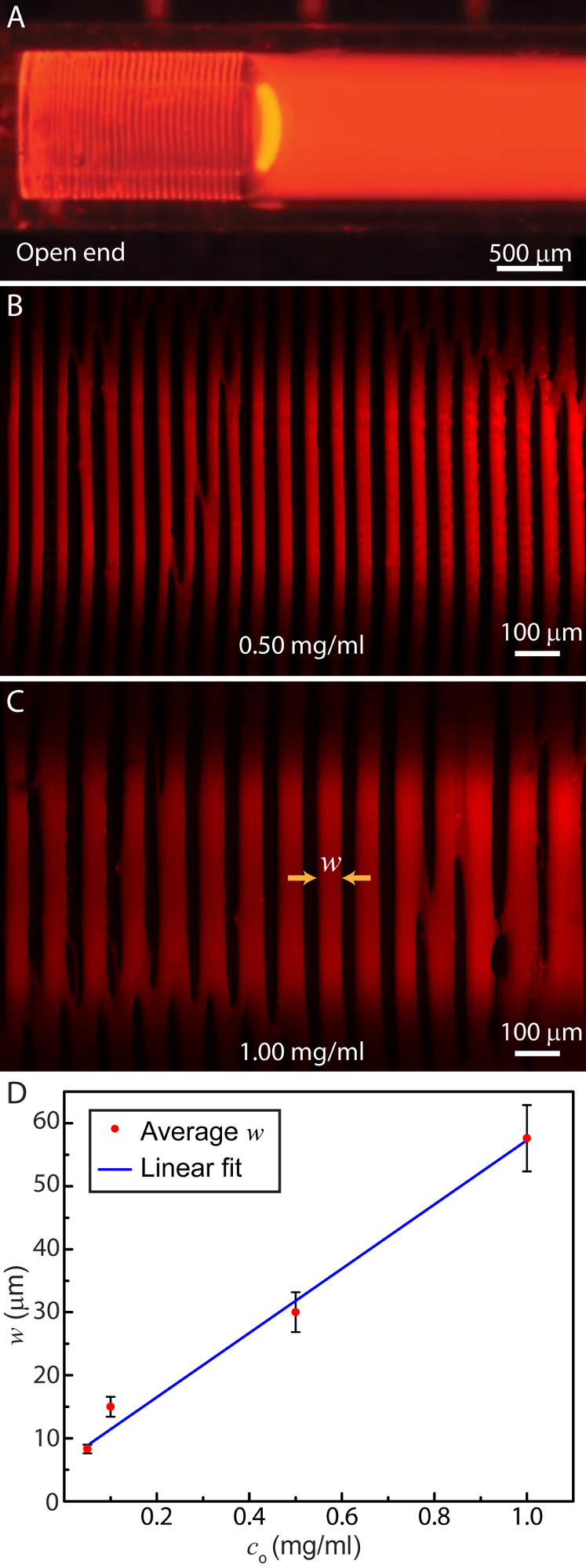}
\caption{QDs bands inside a capillary tube and the concentration dependence of the band width. \textbf{(A)} Image of bands formed from the QDs as toluene evaporates from the open end on the left.  $c_0$ = 0.50 mg/ml. \textbf{(B,C)} Confocal microscopy image showing QD band arrays along the tubes; with initial concentration, $c_0$, of 0.50 mg/ml (0.20 vol\%) and 1.00 mg/ml (0.40 vol\%). We defined $w$ as the width of the band. \textbf{(D)} The graph shows the average width, measured from five different samples, of a band located at $x_m$ = 3 mm.} \label{fc0}
\end{figure}

Near the opening of the tube, we observed that the QDs self-assembled into micrometer wide bands and this feature is regular over a millimeter length scale, similar to the other reports in the literature as discussed in the introduction. This pattern is visible without a microscope, and is particularly clear at higher concentrations of QDs (for instance, 0.50 mg/ml), as shown in the example in Figure \ref{fc0}A.  We have also verified that there is no observable quenching effect throughout the duration of the experiments (see Supporting Information). We then used a fluorescence confocal microscope to further examine the region coated by particles (e.g. Figure \ref{fc0}B,C). The narrow emission spectrum of QDs provides a clear visualization of the coated region; we note that the color tunability of QDs based on the change of core-size would enable the study to be extended to include a mixture QDs of different sizes.

\begin{figure}
\centering
\includegraphics[width=\linewidth]{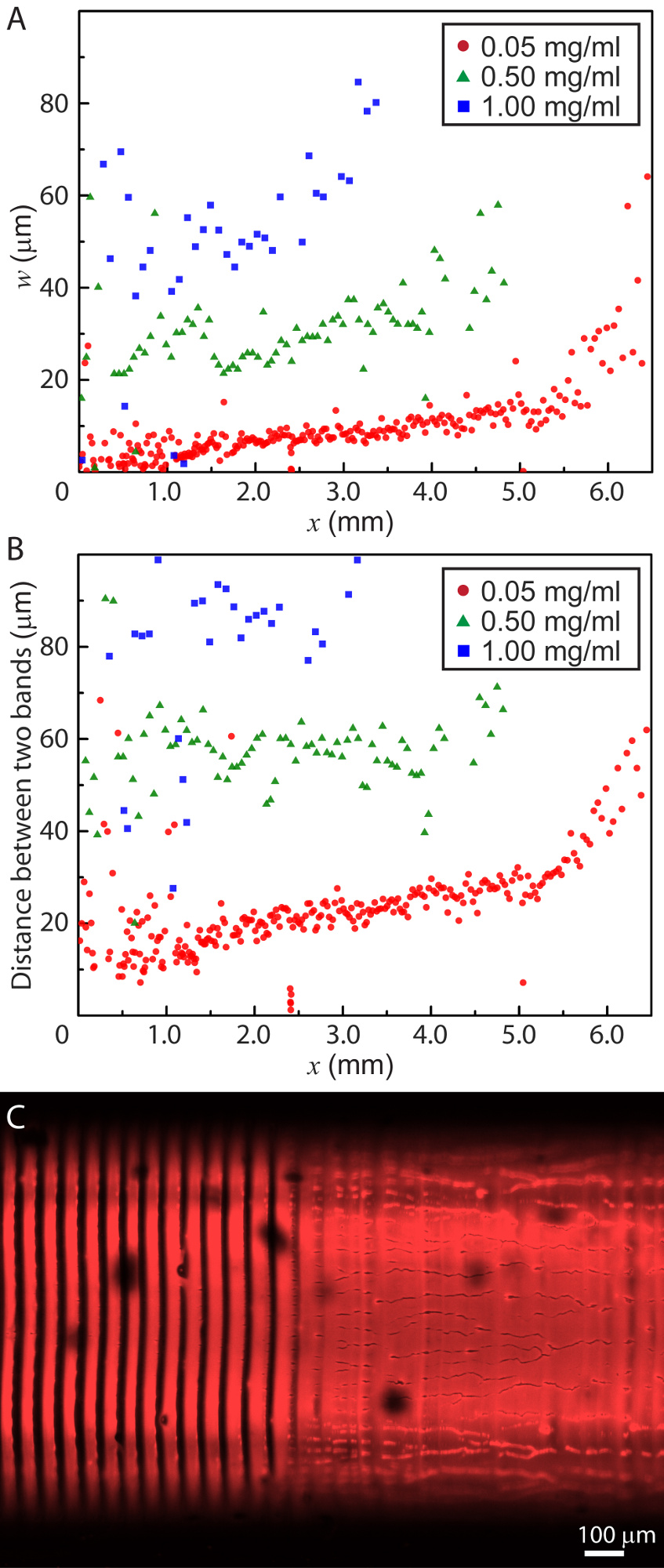}
\caption{QD bands as a function of distance from the tube's opening. \textbf{(A)} The width of the band, \textit{w}, and \textbf{(B)} the distance between two bands measured along the capillary tube for three different initial concentrations, $c_0$. \textbf{(C)} Confocal image shows the transition from bands to a fully coated region ($c_0$ = 0.50 mg/ml, $d$ = 0.90 mm). On the fully coated region, the patterns perpendicular to the band direction are formed from the cracking of the deposited QDs. 
\label{bandx}}
\end{figure}
\begin{figure}
\centering
\includegraphics[width=\linewidth]{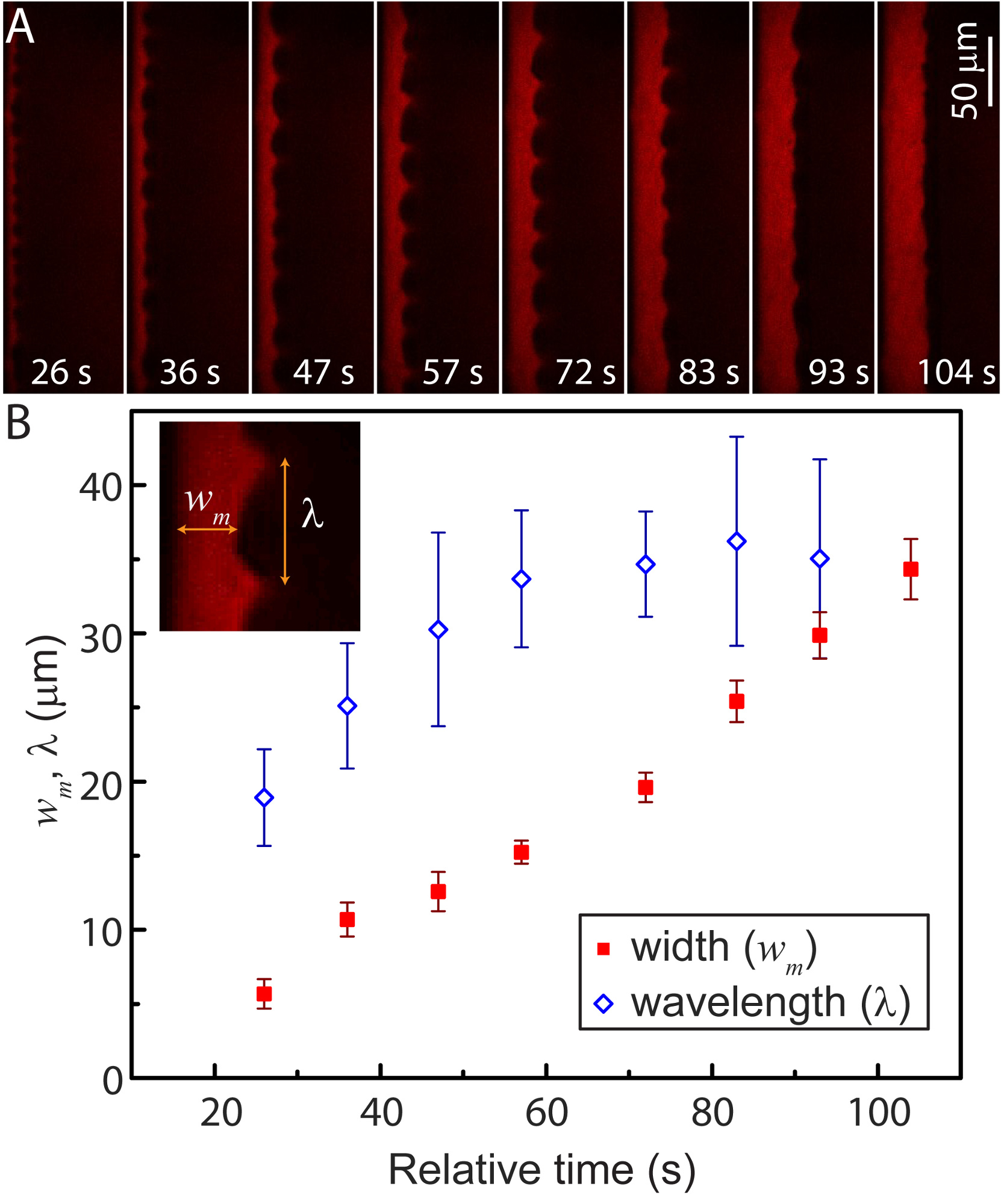}
\caption{Growth of a single band.  \textbf{(A)} Time series images show the growth of a single band ($d = 0.86 \text{ mm}, c_0 = 0.50 \text{ mg/ml}$). \textbf{(B)} The graph shows the relationships between the width, $w_m$, and wavelength, $\lambda$ (see inset for the definition).  
\label{finger}}
\end{figure}
As a means of controlling the width of the band, we examined the dependence of the band width with the initial concentration. To isolate the effect of concentration, we analyzed the width of the bands formed 3 mm away from the tube opening, where the speed of the meniscus movement is approximately 0.4 \si{\um}$\cdot$$\text{s}^{-1}$. We found that the band width increases approximately linearly with an increase of initial concentration, as shown in Figure \ref{fc0}D. This observation is consistent with the increase of the particle flux to the pinned contact line. Larger stripes can be formed at a specific location before the meniscus retracted as more particles are being deposited in the same position in the same amount of time. This result is also consistent with the observations by Watanabe et al. \cite{Watanabe2009} and Kaplan et al. \cite{2014arXiv1412.1813N}  in a planar geometry, where  a dependence of the band width with the concentration of the particles was reported.  

 \begin{figure}
\centering
\includegraphics[width=\linewidth]{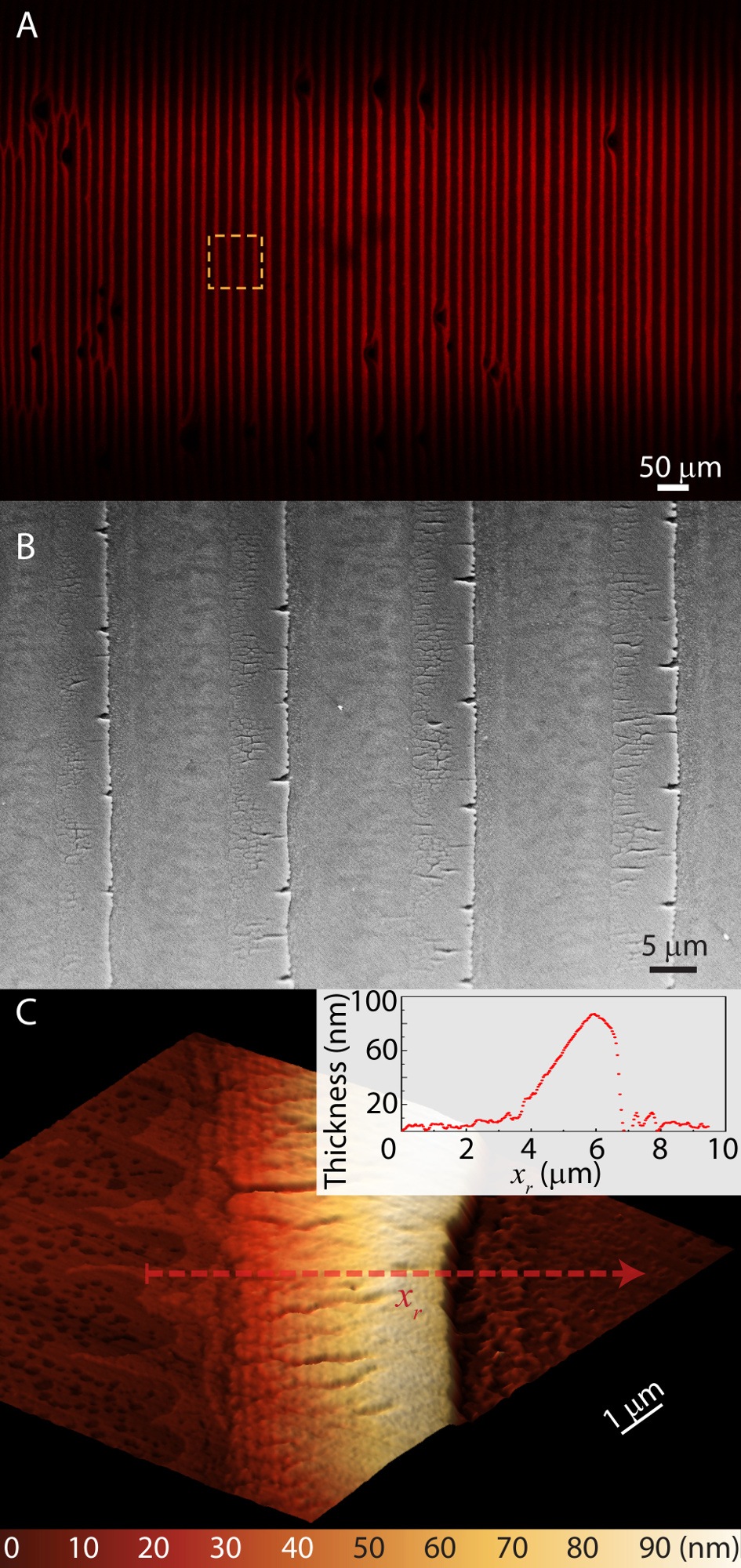}
\caption{QDs bands inside a capillary tube. \textbf{(A)} Confocal microscopy image showing QD band arrays along the tubes; concentration $c_0$ = 0.05 mg/ml (0.02 vol\%). \textbf{(B)} Secondary electron microscopy image (orange box in A) shows the morphology and regularity of the band array, with a band width of 5 \si{\um}. \textbf{(C)} Atomic force microscopy image of one of the bands; color represents the thickness of the QD film. Inset shows the profilometry measurement extracted along the path ($x_r$) of the red arrow on the AFM image. The band thickness increases linearly as the meniscus travels with $x_r$ (4 \si{\um} to 6 \si{\um}), before a sharp decrease as the meniscus slipped to form the next band.\label{SEM}}
\end{figure}
These particulate bands have also been observed in other studies with a confined cylindrical configuration with polystyrene beads \cite{Abkarian2004}, tobacco mosaic virus \cite{Lin2010a}, proteins \cite{Lin2010b} and polymers \cite{sun2015self}. However, it is important to note that the origin of such formation also differs based on the experimental configuration. For instance, in Abkarian et al.'s experimental configuration  \cite{Abkarian2004}, the bands are formed when the contact lines are depinned due to the relative increase of column height as liquid evaporates faster from the bath. In our experimental configuration, the band formation of the meniscus is only affected by the evaporation rate at the tube opening and the local concentration of the QDs as the meniscus is controlled by the evaporation of the toluene, and only one end of the capillary tube is open. This configuration enables us to understand the key factors that determines the features of the band formation.

Next, we examined the dependence of the band structures on the distance from the tube opening.  We found that in general, the width of the band increases as the distance from the opening increases, as shown in Figure \ref{bandx}A. This observation could be affected by two factors: first, the contact line speed decreases as the distance from the tube open end ($x_m$) increases, as described in Figure \ref{Fig2}. Secondly, the concentration of the suspension increases as toluene evaporates. The quantitative relationships between these factors and the deposited patterns will be described in a future study. We have also examined the distance between two bands and its relationship with the location of the tube, $x$. The distance is measured between the two center points of the bands. It is found that with the exception of the 0.05 mg/ml concentration, most of the spacing between the two bands remain approximately constant. As the width grows to be larger than the spacing between the bands, the bands overlap and form a fully coated film, as shown in Figure  \ref{bandx}C.

A closer inspection of the confocal images of the bands shows a subtle undulation at the right side of the band (for instance, at the bands of Figure \ref{fc0}B). Similar observations can also be seen in the bands reported in the literature \cite{Lin2010}. However, the growth of the fingers of the band are not described in detail, partly due to the challenge of visualizing the deposition process in real-time without fluorescence microscopy. In our experiment, the high photoluminescence efficiency of quantum dots enables the direct observation of the band growth with a fluorescence confocal microscope (see Supporting Information, Movie 1). A time series of images shows the growth of a single band in Figure \ref{finger}A. Fingers can be seen during the growth and the pattern moves laterally along the band as it grows. In most cases, the fingers are smoothed before the meniscus front has slipped. In general, the width grows approximately linearly with time while the wavelength increases in the beginning of the growth but saturates at the end, as shown in the graph in Figure \ref{finger}B. The plotted values represent the average of the minimum width per period, $w_m$, and  average wavelength, $\lambda$, (see inset at Figure \ref{finger}B for the definition) that are observed in the time series images during the growth of a single band as shown in Figure \ref{finger}A.

To examine the morphology of the QD bands, the capillary tube is cleaved and examined with a scanning electron microscope (SEM). An array of regular bands with 5 \si{\um} width is observed (Figure \ref{SEM}B), as well as some dewetted tails in the gap between the bands. SEM images also reveal cracks on the bands. The origin of cracks may originate from the contraction of the QDs through capillary pressure, which is being constrained by the adhesion of the QD film with the glass substrate \cite{Groisman1994,Chiu1993}. Further, SEM images also show a smaller crack wavelength at the beginning of the band than the end of the band, which suggests that the deposit thickness increases in the direction of the drying \cite{Allain1995}.

Indeed, examining one of the QD bands with atomic force microscopy verified the above observation. Profilometry data shows that the deposit thickness increases linearly as the band grows to a relative thickness of 90 nm before a sharp decrease of thickness (Figure \ref{SEM}C). This thickness profile suggests that the band grows gradually when particles accumulate at the pinned contact line, before the meniscus slips to the next position as the toluene evaporates. This observation is also consistent with reports in the literature, which used spherical polystyrene particles \cite{Abkarian2004,Bodiguel2010} and a similar thickness profile of the bands was measured. On the other hand, the thickness profile with Tobacco Mosaic Virus rods \cite{Lin2010a} (300 nm $\times$ 18 nm) is observed to generate a band structure with monolayer thickness. The differences could be due to the aspect ratio of the particles, which affects the packing structure when the particles accumulate at the pinned contact line.

\section{Conclusion}
We presented a method to study the deposition of functional nanoparticles in a confined geometry. The intrinsic photoluminescence  property of QDs enables clear observations of the coating processes without the interference from fluorescent polymers.  The experimental platform also enables the real-time measurement of the evaporation speed as well as the visualization of the growth of bands, even at a low volume fraction of QDs. Decoupling the evaporation rate and the concentration in the liquid phase requires further investigations. For instance, the direct measurement of concentration is possible by calibrating the volume fraction of QDs with its photoluminescence intensity.  These understandings could provide key insights on the coating of nanoparticles in a confined geometry and ultimately enable the creation of functional devices inside such constructs.  

\paragraph{acknowledgement}
The authors thank Dr. Sangwoo Shin, Jesse Ault, Dr. Jason Wexler  and Prof. Michael C. McAlpine  for helpful discussions throughout the project. Y.L.K. thanks the Department of Mechanical and Aerospace Engineering, Princeton University for partial support. F.B. acknowledges that the research leading to these results received funding from the People Programme (Marie Curie Actions) of the European Union's Seventh Framework Programme (FP7/2007-2013) under REA grant agreement 623541.

%%%%%%%%%%%%%%%%%%%%%%%%%%%%%%%%%%%%%%%%%%%%%%%%%%%%%%%%%%%%%%%%%%%%%
%% The same is true for Supporting Information,f which should use the

    \bibliography{bands}
    \bibliographystyle{unsrt}

    \end{document}